\documentstyle[preprint,prd,aps,eqsecnum,psfig]{revtex}\tighten

\newcommand{\beq}{\begin{equation}}
\newcommand{\eeq}{\end{equation}}
\newcommand{\beqa}{\begin{eqnarray}}
\newcommand{\eeqa}{\end{eqnarray}}
\newcommand{\simg}{\gtrsim}

\newcommand{\p}{\phi}
\newcommand{\dphi}{\dot\phi}
\newcommand{\k}{\kappa}

\begin{document}

\draft
\preprint{\tighten\vbox{\hbox{UTAP-352}\hbox{astro-ph/9912463}}}

\title{Kinetically Driven Quintessence
}

\author{
Takeshi Chiba\footnote{
Electronic address: chiba@utap.phys.s.u-tokyo.ac.jp, JSPS research
fellow.}, 
Takahiro Okabe\footnote{
Electronic address: okabe@utap.phys.s.u-tokyo.ac.jp
}, 
and Masahide Yamaguchi\footnote{
Electronic address: gucci@utap.phys.s.u-tokyo.ac.jp, JSPS research
fellow.}
}

\address{
Department of Physics, University of Tokyo, 
Tokyo 113-0033, Japan
}

\date{November 12, 1999}

\maketitle

\begin{abstract}
Recently, a novel class of models for inflation has been found in
which the inflationary dynamics is driven solely by (non-canonical)
kinetic terms rather than by  potential terms.
As an obvious extension, we show that a scalar field with
non-canonical kinetic terms alone behaves like an energy component
which is time-varying and has negative pressure presently, {\it i.e.}
quintessence.
We present a model which has a constant equation of state, that is, a
``kinetic'' counterpart of the Ratra-Peebles model of a quintessence
field with a potential term.
We make clear the structure of the phase plane and show that the
quintessential solution is a late-time attractor.
We also give a model for the ``phantom'' component which has an equation
of state with $w=p/\rho <-1$.

\end{abstract}

\section{introduction}\label{introduction}

Recent indirect or direct observations suggest that the Universe
is currently dominated by an energy component with negative
pressure\cite{os,kt,sn1,sn2}. 
One possibility for such a component
is the cosmological constant. Another possibility is dynamical
vacuum energy or quintessence, a temporary decreasing and spatially
inhomogeneous component with negative 
pressure\cite{dolgov,ozer,rp,fn,fhw,csn,tw,cds,zws,bs,chiba,matarrese,masiero,exponential}. 
Only recently, a more radical candidate (called ``phantom'' component)
has been proposed which is ``growing'' in time\cite{caldwell}.

Two problems arise from such a vacuum energy.
The first is the fine-tuning problem: the vacuum energy density of
order $\sim10^{-47}{\rm GeV}^4$ requires the introduction of a new
mass scale about 14 orders of magnitude smaller than the electroweak
scale.
The second is the coincidence problem: the conditions in the early
universe have to be set very carefully in order for the energy density
of the vacuum and that of the matter to be comparable today.
These problems are degenerate for the cosmological constant, however,
they are separated in quintessence.
A class of quintessence can avoid the coincidence problem by means of
the attractor solution\cite{rp,zws}.
It is shown that the quintessence field approaches a common
evolutionary track for a very wide range of initial conditions, so
that the cosmology is extremely insensitive to the initial conditions.

Usually the quintessence field is modeled by a scalar field with a
canonical kinetic term and a potential term.
However, we show that a scalar field with solely kinetic terms can 
(even without potential terms), albeit they are non-canonical, mimic
such a (canonical) quintessence field. Our model is a natural
extension of the kinetically driven inflation model proposed 
recently\cite{adm}.
A mechanism is proposed by which a dilaton remains massless\cite{dp}. 
Assuming universality of the dilaton coupling functions, it has been
shown that the dilaton evolves cosmologically towards values where it 
decouples from matter (so called ``Least Coupling
Principle'')\cite{dp}. 

Unlike the usual potentially driven quintessence model which
automatically satisfies the weak energy condition, 
allowing non-canonical kinetic terms enables us to model
the missing energy component which violates even the weak energy condition
(so called ``phantom field''\cite{caldwell}). 
Only recently, Caldwell draw our attention to consider more general
equation of state with $w=p/\rho<-1$\cite{caldwell}. As he noted, since 
such a ``phantom'' equation-of-state cannot be achieved with a 
canonical Lagrangian and Einstein gravity, considering the phantom field 
requires some extension: either (i) non-canonical Lagrangian or (ii)
non-Einstein  gravity (or both). The latter possibility seems unlikely 
since the deviation from general relativity at the present time is
strongly constrained by the solar system experiments\cite{will}. Our
attempt is minimal one:
non-canonical kinetic terms without a potential term. 
We intend to develop a more general study by including potential terms
as well in the near future. 

The organization of the paper is as follows.
In Sec.\ref{basics}, we present our model.
In Sec.\ref{quintessence}, we start to show the existence of a
scaling solution with a constant equation of state, $-1<w<0$.
Then we show that the scaling solution is a late-time attractor by
means of linear and numerical analyses.
In Sec.\ref{phantom}, we give a model which has a scaling solution 
with a constant equation of state of  $w<-1$, and show that the scaling
solution is a late-time attractor.
In Sec.\ref{reconstruct}, we make a comment on the 
possibility of reconstructing Lagrangian through observational data.
Sec.\ref{summary} is devoted to summary.

\section{basics}\label{basics}

We consider the following action of a single scalar field $\phi$
minimally coupled with gravity:
\beq
S = \int d^4x\sqrt{-g}\left({1\over 2\kappa^2}R+p(\p,\nabla
  \p)\right)+S_B,
\label{eq:action}
\eeq
where $\kappa^2\equiv 8\pi G$ and $S_B$ denotes the action of the
background matter and/or radiation.
Following\cite{adm}, for simplicity, we only consider Lagrangians
which depend only on the scalar field $\phi$ and its derivative
squared\footnote{We use the metric signature  $(-+++)$.}  
\beq
X \equiv -{1\over 2}\nabla^{\mu}\p\nabla_{\mu}\p.
\eeq
For the general action of the scalar field (\ref{eq:action}), the
field equations are given by
\beq
R_{\mu\nu}-{1\over 2}g_{\mu\nu}R =
\kappa^2\left(\frac{\partial p(\p,X)}{\partial X}\nabla_{\mu}\p
\nabla_{\nu}\p+p(\p,X)g_{\mu\nu}+T^B_{\mu\nu}\right)
\label{eq:einstein}
\eeq
where $T^B_{\mu\nu}$ denotes the energy-momentum tensor of the
background.
Eq.(\ref{eq:einstein}) shows that $p(\p,X)$ in the action
(\ref{eq:action}) actually corresponds to the ``pressure'', $p_{\p}$,
of the scalar field\cite{adm}, while the energy density, $\rho_{\p}$,
is given by $\rho_{\p}=2X\partial p/\partial X-p$. 
Hence the extrema of $p(\p,X)$ with respect to $X$ correspond to the
same equation of state as that of a cosmological constant:
$\rho_{\phi}+p_{\phi}=2X\partial p/\partial X=0$.

We assume that the universe is described by a flat homogeneous and
isotropic universe model with the scale factor $a$.
The time coordinate is so normalized that $a=1$ at present.
The field equations are then
\beqa
&&H^2 := \left({\dot a\over a}\right)^2
= \frac{\kappa^2}{3}(\rho_B+\rho_{\p})
= \frac{\kappa^2}{3}\left(\rho_B+2X{\partial p\over\partial
    X}-p\right), \label{eq:hubble} \\
&&{\ddot{a}\over a} = -{\kappa^2\over
  6}(\rho_B+3p_B+\rho_{\p}+3p_{\p}),
\label{eq:a} \\
&&\dot\rho_B = -3H(\rho_B+p_B) =: -3H(1+w_B)\rho_B,
\label{eq:conservation}\\
&&\ddot{\p}\left({\partial p\over \partial X}+
  \dot\p^{2}{\partial^{2}p\over \partial X^2}\right)+3H{\partial
  p\over \partial X}\dot\p+{\partial^{2}p\over \partial X\partial
  \p}\dot\p^{2}-{\partial p\over \partial\p} = 0
\label{eq:q_eom1}
\eeqa
where $\rho_B$ and $p_B$ are the energy density and the pressure of
the background matter and/or radiation, respectively. 

Since we only consider kinetic terms, we must impose that the function 
$p(\phi,X)$ vanishes when $X \rightarrow 0$.\footnote{This amounts to
  assuming some resolution of the cosmological constant problem. The
  situation is the same as assuming the minimum of the 
  potential energy is zero in the (canonical) quintessence field with a
  potential term.}
Near $X=0$, a generic Lagrangian may be expanded as
\beq
p(\phi,X) = K(\phi)X+L(\phi)X^2+\cdot\cdot\cdot.
\label{eq:p_general}
\eeq

\section{power-law kinetic quintessence}\label{quintessence}

To see the effect of non-canonical kinetic terms in a concrete matter, 
in this section, we shall concentrate on the simplest Lagrangian containing
only $\dot{\phi}^2$ and $\dot{\phi}^4$ terms, namely
\beq
p(\phi,X) = K(\phi)X+L(\phi)X^2.
\label{eq:p}
\eeq
In order to realize a model with negative pressure, $K$ and/or $L$
should be negative (note that $X\geq 0$).
However, for the positivity of $\rho_{\p}$ for large $X$, we assume
that $L$ is always positive.
We thus consider the case of $K<0$.
By redefining the scalar field and working with new field variable
such that 
\beq
\phi_{{\rm new}} = \int^{\phi_{\rm old}} d\phi
\frac{L(\phi)^{1/2}}{|K(\phi)|^{1/2}},
\eeq
we rewrite Eq.(\ref{eq:p}) as
\beq
p(\p,X) = f(\p)(-X+X^2),
\label{eq:p_phi}
\eeq
where $\p \equiv \p_{\rm new}$, $X \equiv X_{{\rm new}} =
(L/|K|)X_{{\rm old}}$ and $f(\p) \equiv K^{2}(\p_{{\rm old}})/
L(\p_{{\rm old}})$.
We may regard Eq.(\ref{eq:p_phi}) as the basic
Lagrangian.
Then the pressure, $p_Q$, and energy density, $\rho_Q$, of the
quintessence is given by
\beqa
p_Q &=& f(\p)(-X+X^2),
\label{eq:p_Q} \\
\rho_Q &=& 2X{\partial p\over \partial X}-p= f(\p)(-X+3X^2).
\label{eq:rho_Q}
\eeqa

\subsection{scaling solution}

We look for scaling solutions which keep $w_Q \equiv p_Q/\rho_Q$
constant.
Then from Eqs.(\ref{eq:p_Q}) and (\ref{eq:rho_Q}), $X$ is also
found to be constant:
\beq
X = \frac{1-w_Q}{1-3w_Q}.
\label{eq:x}
\eeq
During the matter or radiation dominated epoch ($\rho_B\gg \rho_Q$), 
Eq.(\ref{eq:q_eom1}) becomes
\beq
\dot \rho_Q = -{2\over t(1+w_B)}(1+w_{Q})\rho_Q .
\label{eq:q_conservation}
\eeq
Substituting Eqs.(\ref{eq:rho_Q}) and (\ref{eq:x}) into the above 
equation, we thus obtain
\beq
f(\p) \propto (\p-\p_*)^{-2(1+w_Q)/(1+w_B)},
\label{eq:f}
\eeq
where $\p_*$ is a constant.
For simplicity, we henceforth choose $\p_* = 0$. 

To summarize, for the scalar field model with the constant equation of
state, $w_Q$, during the matter or radiation dominated epoch, the
function $f(\p)$ should take the form of Eq.(\ref{eq:f}).
Conversely, if the function $f(\p)$ is given by
\beq
f(\p) \propto \p^{-\alpha},
\label{eq:f_q}
\eeq
then there exists a scaling solution such that the equation of state
is characterized by 
\beq
w_Q = \frac{(1+w_B)\alpha}{2}-1.
\label{eq:eos_q}
\eeq
Hence if we require that $w_Q<0$ during the matter dominated epoch,
then the exponent $\alpha$ should satisfy
\beq
\alpha < 2.
\eeq

Note that the weak energy
condition ($w_{Q}\geq -1$) can be violated if $\alpha <0$. 
The stability against perturbations is signified by the ``speed of
sound'' defined by\cite{adm}
\beq
c_s^{2}={p_{Q,X}\over \rho_{Q,X}}={p_{,X}\over p_{,X}+2Xp_{,XX}}.
\label{eq:sound}
\eeq
For  the quartic model Eq.(\ref{eq:p_Q}), one finds
$c_{s}^{2}=(1+w_Q)/(5-3w_Q)$, and the model is unstable for
perturbations on all length scales if the weak energy condition is
violated: $w_Q<-1$.
One may wonder this is always the case.
However, this is not so. 
In fact, we can consider a more general Lagrangian of the
form\cite{adm}
\beq
p(\p,X)=f(\p)g(X),
\label{eq:general}
\eeq
where $g(X)$ is an arbitrary function of $X$.
One can show the equation of motion Eq.(\ref{eq:q_conservation}) has a 
solution $X_{0}={\rm const.}$ with the function $f(\p)$ being the same 
form as Eq.(\ref{eq:f}) if $X_{0}$ satisfies
\beq
2{\partial \ln g\over \partial \ln X}\bigg{|}_{X=X_{0}}= {1+w_{Q}\over 
  w_{Q}}.
\eeq
Note that Eq.(\ref{eq:sound}) involves the second derivative 
$g_{,XX}(X_{0})$.
Therefore, one can always arrange it so that $c_{s}^{2}>0$ for
stability. 
This may be a concrete realization of the ``phantom'' field (or
``growing lambda'') using the
non-canonical Lagrangian\cite{caldwell}. An example of such a phantom
field will be considered in Sec.\ref{phantom}. In this section, we
shall limit ourselves to the case of $0<\alpha <2$ so that $-1<w_Q<0$.

\subsection{attractor structure}

In the matter or radiation dominated universe, there exists the
scaling solution for the kinetic Lagrangian.
Here, we show that this solution is an attractor of the equation of
motion for the scalar field.
When $\rho_B\gg \rho_Q$, Eq.(\ref{eq:q_eom1}) becomes
\beq
\ddot{\p}(1-3\dot{\p}^2)+\frac{2}{t(1+w_B)}(1-\dot{\p}^2)\dot\p+{f'\over
  4f}\left(2-3\dot{\p}^2\right)\dot\p^{2} = 0.
\label{eq:q_eom2}
\eeq
Since this equation has reflection symmetry $\p\leftrightarrow -\p$,
we mainly consider the case of $\p>0$.

Following\cite{rp,ls}, we make the change of variables as
\beqa
\tau &\equiv& \ln t,
\label{eq:tau} \\
u &\equiv& \frac{\p}{\p_s},
\eeqa
where the scaling solution, $\p_s$, is given by
\beq
\p_s = \sqrt{\frac{2(1-w_Q)}{1-3w_Q}}\ t =: \xi_s t.
\label{eq:scaling}
\eeq
With these changes, Eq.(\ref{eq:q_eom2}) becomes
\beqa
u' &=& v \label{eq:u_eom}, \\
v' &=& - v + \frac{1}{1-3\xi^2_s(v+u)^2} \nonumber \\
& &\times\left[\frac{2}{1+w_B}\{-(v+u)+\xi^2_s(v+u)^3\}
-\frac{\alpha}{4u}\{-2(v+u)^2+3\xi^2_s(v+u)^4\}\right],
\label{eq:v_eom}
\eeqa
where $'$ denotes the derivative with respect to $\tau$.
Then one can find three critical points; $(u,v)=(0,0),(1,0)$ and
$(-1,0)$.
The $(-1,0)$ critical point corresponds to the scaling solution with 
negative amplitude, while the $(0,0)$ critical point is a trivial one
in which $X=0$. 

In order to study the stability near the critical points, we use
linear analysis\cite{rp,ls}.
Perturbing about the scaling solution $(u,v)=(1+\delta u,0+\delta v)$
and keeping only the terms linear in $\delta u$ and $\delta
v$, Eqs.(\ref{eq:u_eom}) and (\ref{eq:v_eom}) become
\beqa
\delta u' &=& \delta v, \\
\delta v' &=& -\delta v+
\frac{2w_Q}{1+w_B}(\delta u+\delta v)-\frac{1+w_Q}{2(1+w_B)}
\frac{2-3\xi^2_s}{1-3\xi^2_s}\delta u.
\eeqa
Then the eigenvalues of small perturbations are given by
\beq
\lambda^{\pm}_s = \frac{2w_Q-1-w_B\pm
\sqrt{(2w_Q-1-w_B)^2+8(1+w_B)\{w_Q-(1+w_Q)/(5-3w_Q)\}}}{2(1+w_B)}.
\eeq
The necessary and sufficient condition for stability is that the real
part of the eigenvalues be negative.
Note that because $-1<w_Q<0$, the second term under the square root is
always negative.
Then the condition for the stability is just $2w_Q-1-w_B<0$.
Hence the $(1,0)$ critical point corresponding to the scaling solution
is stable.
In a similar way, the eigenvalues of small perturbations near the
trivial solution is given by
\beq
\lambda^{\pm}_t =
\frac{2w_Q-1-w_B\pm\sqrt{(2w_Q-1-w_B)^2+4(1+w_B)(w_Q-1)}}{2(1+w_B)}.
\label{eq:lambda_tr}
\eeq
Then the $(0,0)$ critical point is also stable.

\subsection{numerical analysis}

\begin{figure}

\centerline{\psfig{file=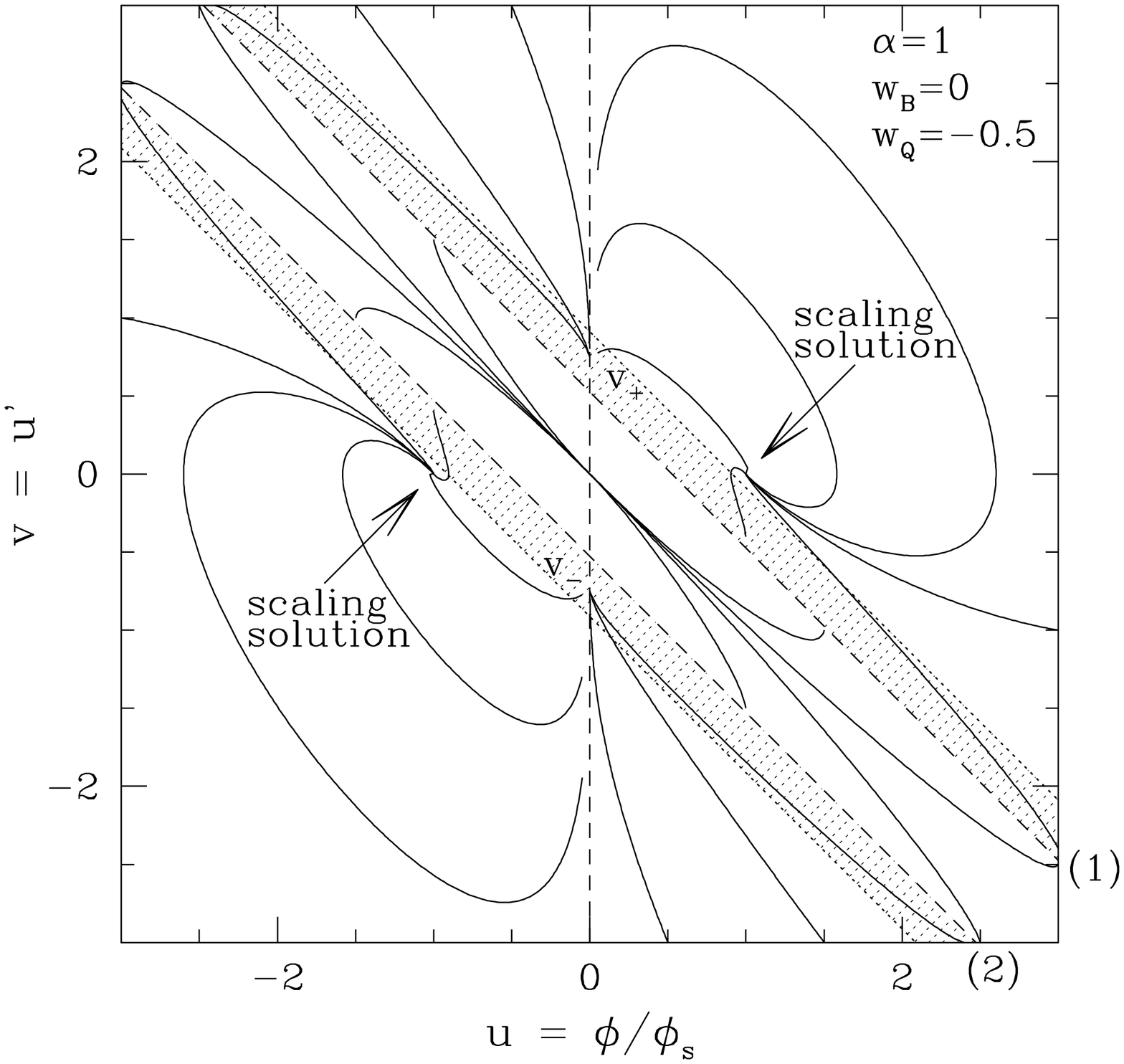,width=\columnwidth}}
\caption{Phase plane of the quintessence field in the matter dominated
  universe. Trajectories in the region between line (1) and (2) converge
  on the trivial solution $(u,v)=(0,0)$. In the region above line (1)
  and $u>0$ they converge on the scaling solution $(1,0)$ directly,
  while in the region below line (2) and $u>0$ they approach and pass
  through the point $(0,v_-)$, then converge on another scaling
  solution $(-1,0)$. The speed of sound is imaginary in the shaded
  region.}
\label{fig:q_phase_space}

\centerline{\psfig{file=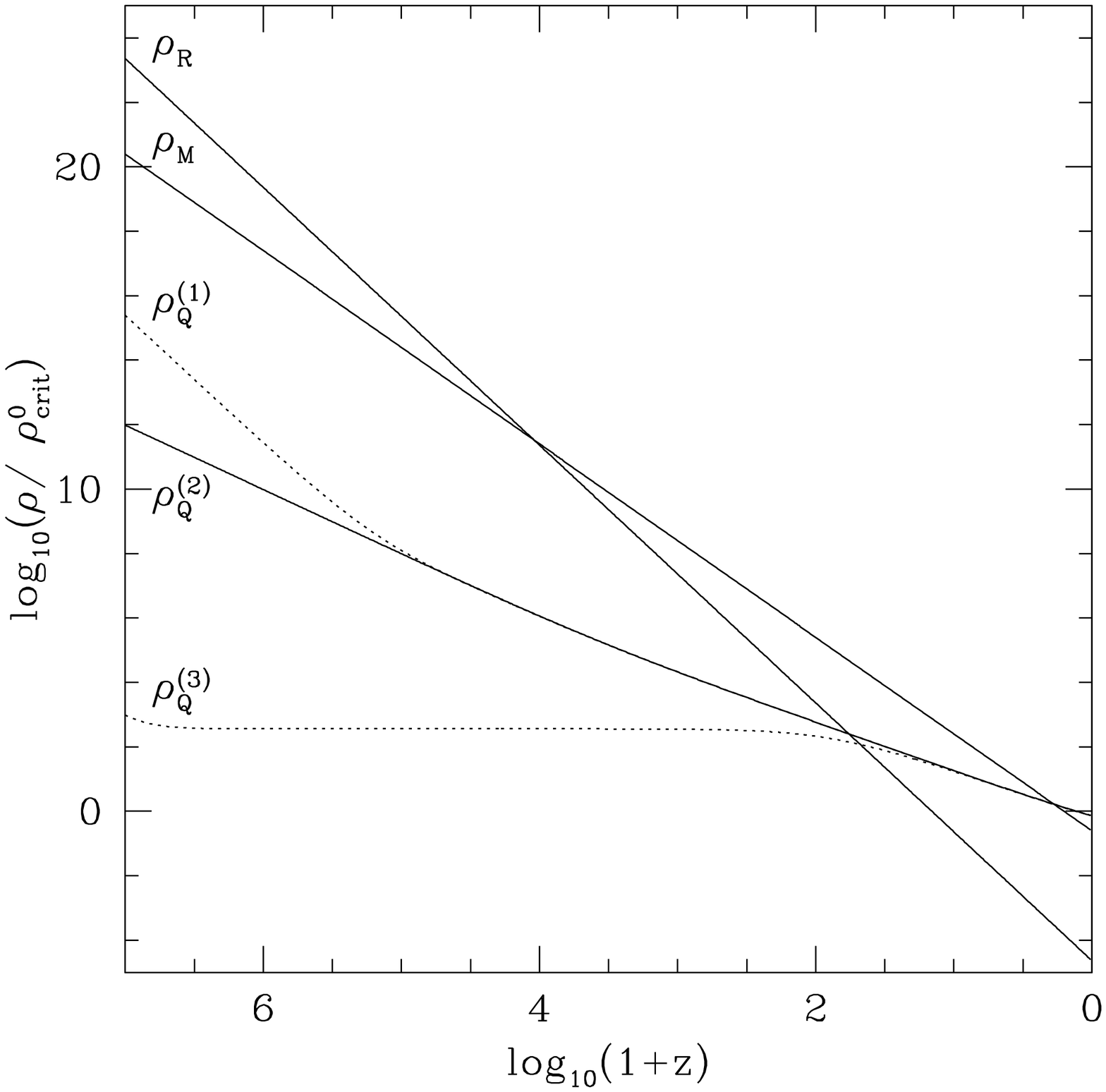,width=\columnwidth}}
\caption{Energy densities of radiation, $\rho_R$, matter, $\rho_M$,
  and quintessence field for various initial conditions,
  $\rho^{(i)}_Q,\ i=1,2,3$, against the redshift. The initial
  conditions for $\rho^{(2)}_Q$ correspond to that of the scaling
  solution in the radiation dominated universe. The initial energy
  density of the quintessence field which converges on the attractor
  solution spans more than 10 orders of magnitude.}
\label{fig:q_density}

\centerline{\psfig{file=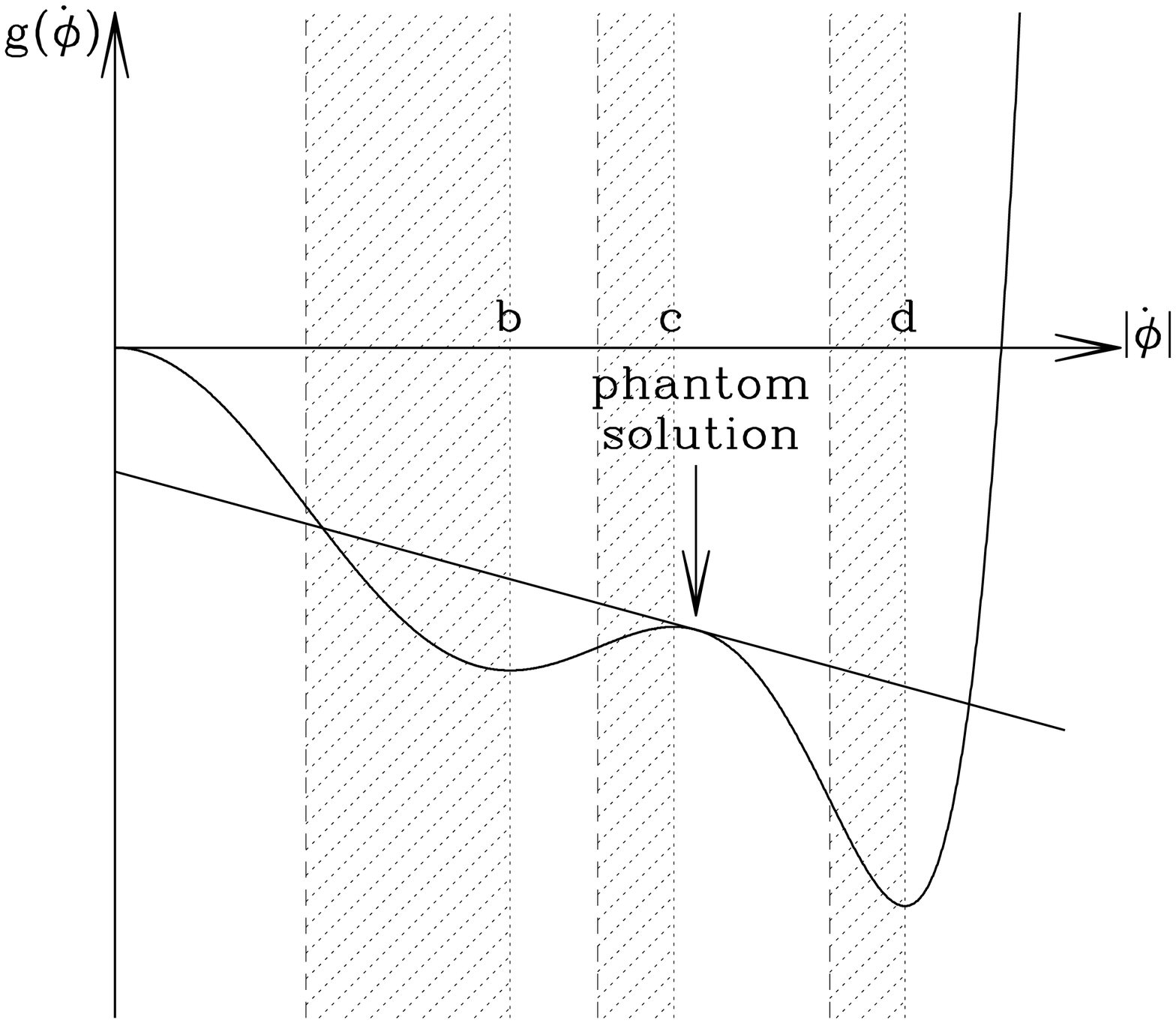,width=\columnwidth}}
\caption{A sketch of the function $g(\dot{\p})$. $c^2_s<0$ in the
  shaded regions. The phantom solution is one of the solutions of
  Eq.(\ref{eq:p_scaling}) at which $\dot{\p}g_{,\dot{\p}}<0$ and
  $g_{,\dot{\p}}/\dot{\p}g_{,\dot{\p}\dot{\p}}>0$ and the intersection
  of the tangent with the vertical axis is negative.}
\label{fig:g}

\centerline{\psfig{file=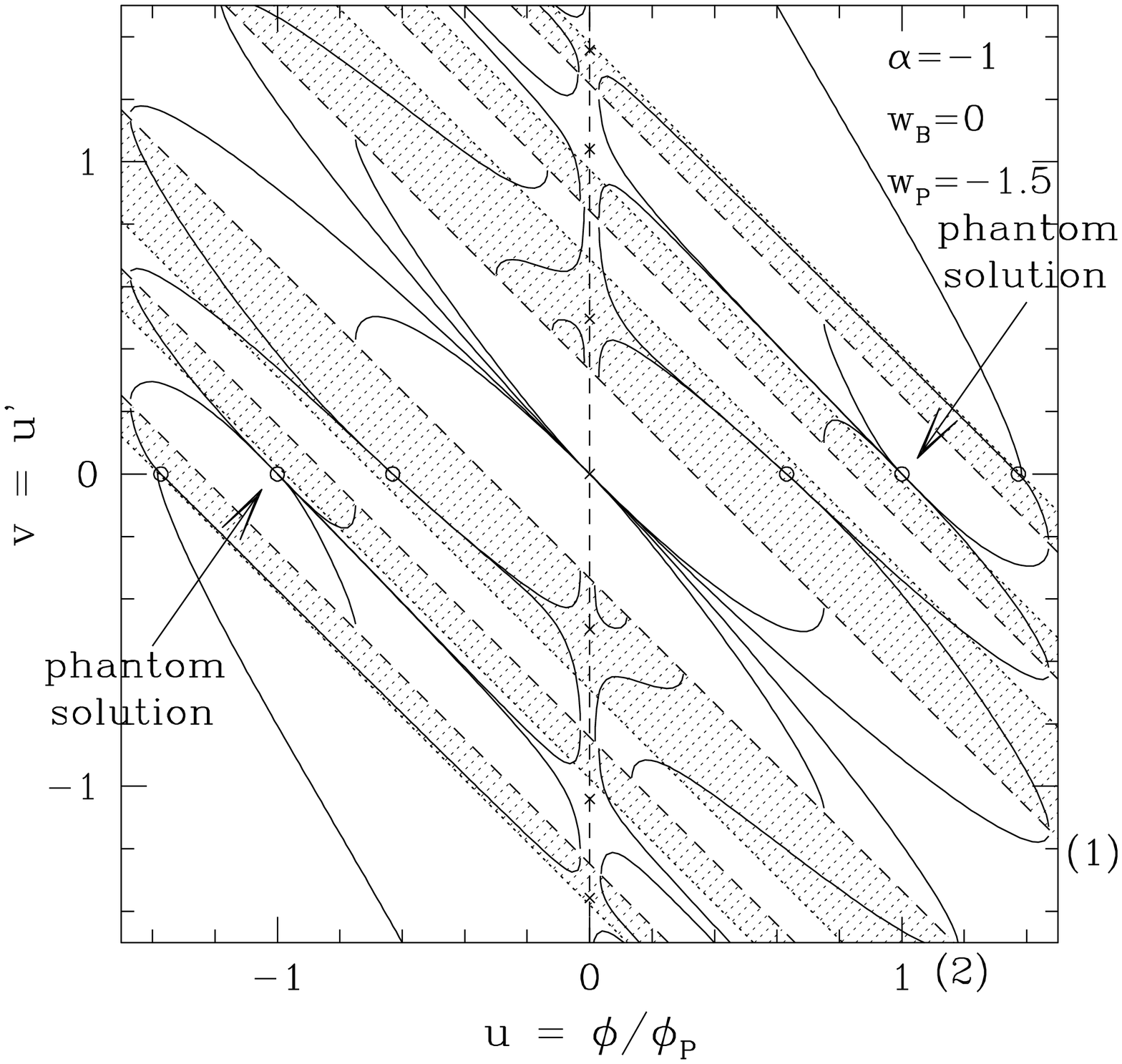,width=\columnwidth}}
\caption{Phase plane of the phantom field in the matter dominated
  universe. We set $b^2=1/2$, $c^2=1$ and $d^2=2$. The scaling
  solutions are represented by open circle. Eq.(\ref{eq:p_v_eom}) is
  singular on the dashed lines except for the points represented by
  cross. In the shaded regions corresponding to those in
  FIG.\ref{fig:g}, $c^2_s<0$. Trajectories in the region between line
  (1) and (2) converge on the trivial solution. In the region above
  line (1) and $u>0$, they converge on the scaling solutions. In the
  region below line (2) and $u>0$, trajectory approaches a point on
  the boundary asymptotically.}
\label{fig:p_phase_space}

\centerline{\psfig{file=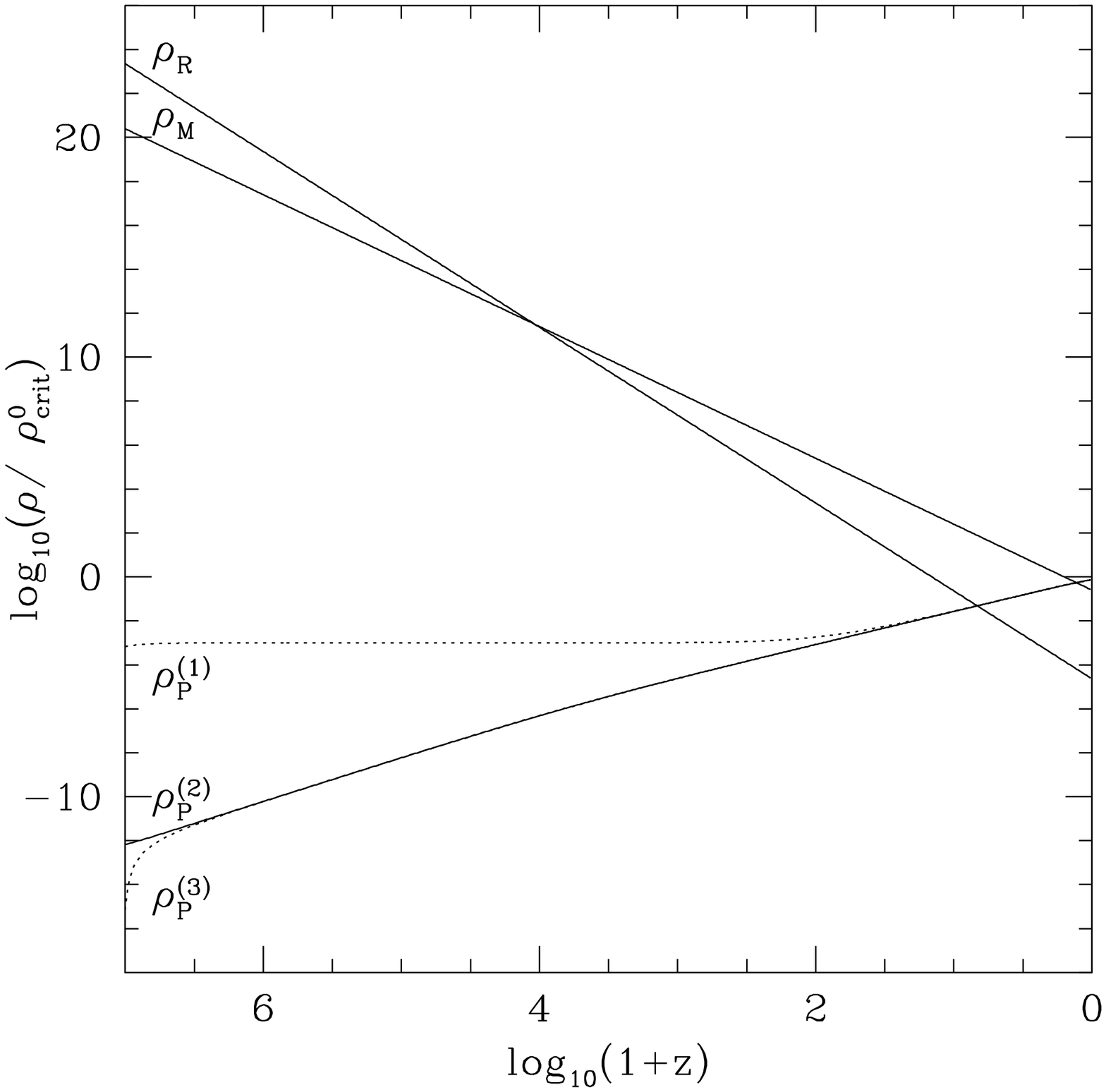,width=\columnwidth}}
\caption{Energy densities of radiation, $\rho_R$, matter, $\rho_M$,
  and phantom field for various initial conditions,
  $\rho^{(i)}_P,\ i=1,2,3$, against the redshift. The initial
  conditions for $\rho^{(2)}_P$ correspond to that of the phantom
  solution in the radiation dominated universe. The initial energy
  density of the phantom field which converges on the phantom
  solution spans more than 10 orders of magnitude.}
\label{fig:p_density}

\end{figure}

We have shown that the scaling solution is stable for small
perturbations.  In order to analyze the phase plane, we solve 
Eqs.(\ref{eq:u_eom}) and (\ref{eq:v_eom}) numerically.
The phase plane is shown in FIG. \ref{fig:q_phase_space} for the case
of $\alpha=1$ and $w_B = 0$.
This figure shows the attractor structure of the scaling solutions and
the trivial solution: there are trajectories which converge on these
solutions asymptotically. The boundaries between phase flow, lines
$(1)$ and $(2)$ in FIG. \ref{fig:q_phase_space}, correspond to the lines 
where Eq.(\ref{eq:v_eom}) is singular:
\beq
v = -u\pm \sqrt{\frac{1-3w_Q}{6(1-w_Q)}}.
\label{eq:singular}
\eeq
Eq.(\ref{eq:v_eom}) is also singular on $u=0$ except for
\beq
v = v_{\pm}:=\pm\sqrt{\frac{1-3w_Q}{3(1-w_Q)}}.
\eeq
The trajectories in the region $v<-u+\sqrt{(1-3w_Q)/6(1-w_Q)}$ and
$u>0$ approach and pass through the point $(u,v)=(0,v_-)$, then
converge on the scaling solution with negative amplitude.
{}From Eqs.(\ref{eq:sound}), the requirement for stability against
perturbations is not satisfied in the region
\beq
\frac{1-3w_Q}{6(1-w_Q)}<(u+v)^2<\frac{1-3w_Q}{2(1-w_Q)}.
\eeq
These regions correspond to the shaded ones in
FIG. \ref{fig:q_phase_space}.

The cosmological evolution of the scalar field is obtained by solving
Eqs.(\ref{eq:a}) and (\ref{eq:q_eom1}) numerically. The initial
conditions for the scale factor are so chosen that the
Friedmann equation (\ref{eq:hubble}) is satisfied. Those 
 for the scalar field are chosen in the region $u>0$ and
$v>-u+\sqrt{(1-3w_Q)/2(1-w_Q)}$ so that $c_s^2 >0$ initially.
In FIG. \ref{fig:q_density}, we show the time evolution of energy
densities of radiation, matter and quintessence field for various
initial conditions. 
The present density parameter of the $i$-th component $\Omega_{i}$ is
defined by $\Omega_{i,0}=\k^{2}\rho^0_i/(3H_{0}^{2})$. 
We choose $\Omega_{M,0}=0.25$ and set $\alpha =1$. 
The figure shows that for a very wide range of initial conditions, the
energy density of the quintessence field converges on a common
evolutionary track.

\subsection{mass scale}
 
 In the Ratra-Peebles model of the quintessence field driven by a
 potential term, one can choose a parameter with mass dimension in the
 potential term to be a typical particle physics scale\cite{zws}. 
 We introduce a parameter with mass dimension in the kinetic Lagrangian
 as
\beq
f(\p) = \frac{M^{4-\alpha}}{\p^{\alpha}},
\label{eq:mass}
\eeq
 then we fix this parameter by requiring that the scalar field is beginning
 to dominate the energy density of the universe today
($\rho^0_{\rm {crit}}$ is the present critical density and $\p_0$ is
 the present value of the scalar field)
\beq
\rho_Q^0 = f(\p_{0})(-X+3X^2) \simeq \rho^0_{\rm crit},
\eeq
 and that the scalar field has already reached the attractor solution
\beq
\left.\frac{f'}{f}\right|_{\p=\p_{0}} \simeq \frac{1}{\p_{0}} \simeq H_0.
\eeq
The last condition is obtained from Eq.(\ref{eq:q_eom2}).
These conditions fix the mass parameter and the present value of the
scalar field as
\beqa
M &\sim& 10^{(43\alpha-48)/(4-\alpha)}\ [{\rm GeV}],
\label{eq:mass_scale} \\
\p_0 &\sim& 10^{43}\ [{\rm GeV}^{-1}],
\eeqa
respectively.
$M$ could be larger than TeV scale when $\alpha \simg 1.3$.

\section{phantom field}\label{phantom}

In this section, we construct a model of a power-law phantom field
which is recently proposed as an alternative to the dark energy
component\cite{caldwell}. In fact, in Ref.\cite{caldwell}, 
a scalar field with a kinetic term of inverted sign is introduced as a 
toy model of such a component: the Lagrangian density is given by ${\cal
L}_P=+{1\over 2}\nabla_{\mu}\phi\nabla^{\mu}\phi -V(\phi)$. The
concern is the tachyonic instability for $\phi$. However, as shown in
\cite{caldwell}, as long as $V_{,\phi\phi}$ is negative, such an 
instability is not developed. 
For example, for a constant equation of state, one can show 
$V_{,\phi\phi}={3\over 2}(1-w_P)[\dot H-{3\over 2}H^2(1+w_P)]$, thus
it is negative as long as $-2<w_P<-1$. 
Our model will not suffer from such a
restriction even for a constant equation of state.

\subsection{scaling solution}

In the previous section, we have mentioned that a general Lagrangian
of the form Eq.(\ref{eq:general}), $p(\phi,X)=f(\phi)g(X)$,  with the 
function $f(\p)$ of the form  Eq.(\ref{eq:f_q}) has scaling 
solutions with $X=$constant.
{}From Eq.(\ref{eq:eos_q}), the equation of state is characterized by
\beq
w_P\equiv\frac{p_P}{\rho_P}=\frac{(1+w_B)\alpha}{2}-1,
\eeq
then the phantom field which violates the weak energy condition,
$w_P<-1$, corresponds to $\alpha<0$.

We determine the function $g(X)$ which has a phantom solution.
We impose the following conditions on $g(X)$ other than $g(X)
\rightarrow 0$ for $X\rightarrow  0$(vacuum triviality) and $g(X) >0$
for $X\rightarrow \infty$(positivity of the energy density for large $X$): 
(i) the positivity of the energy
density $\rho_P=f(2Xg_{,X}-g) >0$ because we are considering the 
missing energy component; 
(ii) the violation of the weak energy condition, $\rho_P +p_P
=2fXg_{,X}<0$; 
(iii) the stability against perturbations, $c_s^2 =g_{,X}/(
g_{,X}+2Xg_{,XX})>0.$ 

In terms of $\dot \phi$ (so that $X=\dphi^2/2$) these conditions are 
rewritten as: 
(i') $g_{,\dphi}\dphi -g >0$, which is geometrically interpreted as 
the intersection of the tangent to the curve $g=g(\dot{\p})$ at the point
$\dot{\p}$ with the vertical axis is negative; 
(ii') $\dphi g_{,\dphi} <0$; (iii') $g_{,\dphi}/(\dphi
g_{,\dphi\dphi}) >0$. Thus, for $\dphi >0$, $g(\dphi)$ should satisfy 
$g_{,\dphi} <0$ and $g_{,\dphi\dphi} <0$.  
Therefore, $g(\dphi)$ must have at least
three extrema for $\dphi>0$ and hence be an eighth-order function of
$\dphi$. 
We parameterize $g(\dphi)$ as\footnote{
We take negative sign for the coefficient of $\dphi^2$ as a
minimal extension of the quartic model Eq.(\ref{eq:p_phi}). If we
take positive sign, then the polynomial should be at least of tenth-order.}:
\beqa
g(\dot{\p})&=& \int^{\dphi}_0\dphi(\dphi^2-b^2)(\dphi^2-c^2)(\dphi^2-d^2) d\dphi\nonumber\\
&=:&
\frac{1}{8}\dot{\p}^8-\frac{1}{6}A\dot{\p}^6+\frac{1}{4}B\dot{\p}^4-
\frac{1}{2}C\dot{\p}^2,
\label{eq:Lagrangian_p}
\eeqa
where $b$, $c$ and $d$ are constants corresponding to the extrema of
$g(\dphi)$ as shown in FIG. \ref{fig:g}. 
A sketch of $g(\dphi)$ which satisfies  the above conditions (i'),
(ii'') and (iii') is shown there. In the shaded regions, $c_s^2 <0$ so 
that the solution is unstable for perturbations on all length scale.
{}From the above equation, the scaling solutions satisfy
\beq
w_P = \frac{C\dot{\p}^2/2-B\dot{\p}^4/4+A\dot{\p}^6/6-\dot{\p}^8/8}
{C\dot{\p}^2/2-3B\dot{\p}^4/4+5A\dot{\p}^6/6-7\dot{\p}^8/8}.
\label{eq:p_scaling}
\eeq
Then there are six scaling solutions at most.

\subsection{attractor structure}

We show that the phantom solution is an attractor of the equation of
motion for the scalar field by means of the linear analysis employed
in the previous section.
{}From Eq.(\ref{eq:Lagrangian_p}), the equation of motion
(\ref{eq:q_eom1}) becomes
\beqa
&&\ddot{\p}(C-3B\dot{\p}^2+5A\dot{\p}^4-7\dot{\p}^6)+\frac{2}{t(1+w_B)}
(C-B\dot{\p}^2+A\dot{\p}^4-\dot{\p}^6)\dot{\p} \nonumber \\
&&+\frac{f'}{f}\left(\frac{1}{2}C\dot{\p}^2-\frac{3}{4}B\dot{\p}^4
+\frac{5}{6}A\dot{\p}^6-\frac{7}{8}\dot{\p}^8\right) = 0,
\label{eq:p_eom}
\eeqa
when $\rho_B\gg\rho_P$.
We make the change of variables as (\ref{eq:tau}) and
\beq
u \equiv \frac{\p}{\p_{P}}=:\frac{\p}{\xi_P t},
\eeq
where $\xi_P$ is determined by one of the solutions of
Eq.(\ref{eq:p_scaling}) corresponding to the phantom solution.
With these changes, Eq.(\ref{eq:p_eom}) becomes
\beqa
u' &=& v,
\label{eq:p_u_eom} \\
v' &=& - v + \frac{1}{C-3B\xi^2_P(v+u)^2+5A\xi^4_P(v+u)^4-7\xi^6_P(v+u)^6}
\nonumber \\
&&\times\left[
\frac{2}{1+w_B}\{-C(v+u)+B\xi^2_P(v+u)^3-A\xi^4_P(v+u)^5+\xi^6_P(v+u)^7\}
\right.
\nonumber \\
&&-\left.\frac{\alpha}{u}\{-\frac{1}{2}C(v+u)^2+\frac{3}{4}B\xi_P^2(v+u)^4
-\frac{5}{6}A\xi^4_P(v+u)^6+\frac{7}{8}\xi^6_P(v+u)^8\}\right],
\label{eq:p_v_eom}
\eeqa
where $'$ denotes the derivative with respect to $\tau$.
Then one can find seven critical points at most; $(u,v)=(0,0),(\pm
1,0)$ and others.

Perturbing about the phantom solution $(u,v)=(1+\delta u,0+\delta v)$
and keeping only the terms linear in $\delta u$ and $\delta v$,
Eqs.(\ref{eq:p_u_eom}) and (\ref{eq:p_v_eom}) become
\beqa
\delta u' &=& \delta v, \\
\delta v' &=& -\delta v+
\frac{2w_P}{1+w_B}(\delta u+\delta v)-\frac{2(1+w_P)}{1+w_B}S\delta u,
\eeqa
where
\beqa
S
&=&\frac{C/2-3B\xi^2_P/4+5A\xi^4_P/6-7\xi^6_p/8}
{C-3B\xi^2_P+5A\xi^4_P-7\xi^6_p} \\
&=&\left.\frac{1}{\xi^2_P}\frac{\dphi g_{,\dphi}-g}{g_{,\dot{\p}\dot{\p}}}
\right|_{\dot{\p}=\xi_P}.
\eeqa
Then the eigenvalues of small perturbations are given by
\beq
\lambda^{\pm}_P = \frac{2w_Q-1-w_B\pm
\sqrt{(2w_P-1-w_B)^2+8(1+w_B)\{w_P-(1+w_P)S\}}}{2(1+w_B)}.
\eeq
Since we impose $\rho_P>0$ and $g_{,\dot{\p}\dot{\p}}<0$ on the
phantom solution, $S<0$.
Then the second term under the square root is negative.
Therefore the $(1,0)$ critical point corresponding to the phantom
solution is stable.
In a similar way, the eigenvalues of small perturbations near the
trivial solution $(0,0)$ is given by Eq.(\ref{eq:lambda_tr}) but $w_Q$
is replaced with $w_P$ since the phase structure around the trivial
solution is determined by the lowest kinetic term.
The trivial solution is also stable.

\subsection{numerical analysis}

In FIG. \ref{fig:p_phase_space}, the phase plane for the case of
$\alpha=-1$ and $w_B=0$ is shown by solving Eqs.(\ref{eq:p_u_eom})
and (\ref{eq:p_v_eom}) numerically. We choose $b^2=1/2, c^2=1$ and
$d^2=2$(or $A=7/2,B=7/2,C=1$). 
There are one trivial solution and six scaling solutions.
The shaded regions correspond to those in FIG. \ref{fig:g},
$c^2_s<0$.
Eq.(\ref{eq:p_v_eom}) is singular on the dashed lines except for the
points represented by cross.
The boundaries between phase flow correspond to these singular lines.
The trajectories in the region above line (1) and $u>0$ converge on
the scaling solutions.
In the region between line (1) and (2), they converge on the
trivial solution.
In the region below line (2) and $u>0$, trajectory approaches a
point on the boundary asymptotically.

The cosmological evolution of the scalar field is obtained by solving
Eqs.(\ref{eq:a}) and (\ref{eq:q_eom1}) numerically.
In FIG. \ref{fig:p_density}, we show the time evolution of energy
densities of radiation, matter and phantom field for various initial
conditions. We choose $\Omega_{M,0}=0.25$ and set $\alpha=-1$.
Since the scaling solution is growing in time and bounded by the
instability regions (in which $w>-1$) as shown in FIG. \ref{fig:g}, 
the initial energy density should be chosen to be below the
present energy scale.\footnote{It is even possible to start from zero
energy density with a fine-tuning of the initial conditions.} 
Hence, there remains a severe fine-tuning
problem of the initial conditions. 
We expect that including potential terms or introducing additional
fields may allow initially decaying solutions. We note, however, that
the condition (ii), $\rho+p=2fXg_{,X}<0$,  remains unchanged even if
we include potential terms. 

\section{reconstructing $p(\p,X)$}\label{reconstruct}

It has been shown that the effective potential of a scalar field with
a canonical kinetic term can be determined by using the
magnitude-redshift relation of distant type Ia
supernovae\cite{cn,alex}.
In a similar way, we consider the possibility of reconstructing $p$ as 
a function of $\phi$  and $X$ through observational data. 
The pressure and density of the quintessence or phantom component can
be written in terms of the coordinate distance to redshift $z$, $r(z)$
(which is related to the luminosity distance, $d_{L}(z)=(1+z)r(z)$) as
\beqa
&&\k^{2}p_{\p}=\k^{2}p(\p,X)=-{3\over
  \left(dr/dz\right)^{2}}-2(1+z){d^{2}r/dz^{2}\over
  \left(dr/dz\right)^{3}},
\label{reconst:p}\\
&&\k^{2}\rho_{\p}=\k^{2}\left(2X{\partial p\over \partial X}-p\right)
={3\over \left(dr/dz\right)^{2}}-3H_{0}^{2}\Omega_{M,0}(1+z)^{3},
\label{reconst:rho}
\eeqa
{}From Eq.(\ref{reconst:p}), we find that $p(\p,X)$ is written as a
function of $z$.
In order to reconstruct $p(\p,X)$, we further need to rewrite $\p$ (or
$X=\dot\phi^2/2$) as a function of $z$.
One might expect that this may be done using
Eq.(\ref{reconst:rho}). Indeed, it is possible for the case of the
quintessence field with a canonical kinetic term and a potential term,
where $p(\p,X) = X - V(\p)$ and $\rho(\p,X) = X + V(\p)$. By summing
up Eqs.(\ref{reconst:p}) and (\ref{reconst:rho}), $X$ can be described
as a function of $z$. Integrating $X$ with respect to $z$ enables us to
rewrite $\p$ as a function of $z$. Then, by subtracting
Eq.(\ref{reconst:p}) from Eq.(\ref{reconst:rho}), $V(\p)$ can be
described as a function of $z$, that is, $\p$. However, in our case,
$p(\p,X)$ is an arbitrary function of both $\p$ {\it and} $X$ so that
it is impossible to rewrite $\p$ or $X$ as a function of $z$ without
identifing the combination of $\p$ and $X$ in $p(\p,X)$.\footnote{We
also note that the reconstruction method fails if the scalar field has
a multi component or is non-minimally coupled\cite{chiba,matarrese} to
the curvature.} 

Of cource, once one specifies the functional form of $p(\p,X)$, the 
reconstruction is possible like the case of the
quintessence field with a canonical kinetic term and a potential term.
For example, for the case of our model Eq.(\ref{eq:p_phi}), we can
write $f(\p)$ and $X$ in terms of observable quantities in the
following manner: Take the ratio of Eq.(\ref{reconst:rho}) to
Eq.(\ref{reconst:p}).
Then $X$ is written as a function of $z$.
Integrating $X$ with respect to $z$, $\p$ can be written as a function of
$z$.
Once again using Eq.(\ref{reconst:p}) together with the
results obtained, we can rewrite $f$ as a function of $z$.
Hence we can reconstruct $f$, $f(\p)=f(\p(z))$.

\section{summary}\label{summary}

We have shown that a scalar field with only non-canonical kinetic
terms can, without the help of potential terms, behave like
quintessence or phantom energy component. 

We have  presented a  kinetic counterpart of
the Ratra-Peebles model and investigated the structure of the phase
plane and shown that the quintessential solution is a 
late-time attractor. 
The phase area of the initial conditions for the scalar field which
converge on the quintessential solution is smaller than that of the
Ratra-Peebles model since there are a trivial solution  as well as the 
scaling solutions in the kinetic Lagrangian. However, for very wide 
ranges of initial energy density, these energy
components converge on common evolutionary tracks. 

We have also given a model of a power-law ``phantom''
field with arbitrary $w$ of $w<-1$ which is stable against
perturbation, and  have shown that the
phantom solution is a late-time attractor. 
In our model of the phantom field, the initial conditions have to be
set carefully because there are a number of scaling solutions.
Moreover, the initial energy density should be chosen to be below the
present energy scale because the decaying lambda region
corresponds to the unstable region. Therefore, there remains a
fine-tuning problem of the initial conditions, which is no more
severer than that of the cosmological constant. The situation may be 
improved by including potential terms or by introducing additional fields. 

The violation of the weak energy condition is required to construct
wormholes\cite{wormhole}. Our model of a phantom field may be used to 
explore such a fascinating possibility.

\acknowledgements
We would like to thank Professor Masahiro Kawasaki and Professor
Takashi Nakamura for useful comments. 
T.C. is also grateful to Dr. Ken-ichi Nakao for critical comments. 
T.O. would like to thank Professor Yasushi Suto for encouragement. 
This work was supported in part by JSPS Fellowship
for Young Scientists under grant No.3596~(TC) and No.4558~(MY).


\end{document}